\documentclass[preprint]{aastex}

\received{}
\accepted{}
\slugcomment{To appear in ApJL (accepted 3/10/2000).}

\shorttitle{GRAIN SURVIVAL IN SHOCKS}
\shortauthors{MOURI \& TANIGUCHI}

\begin{document}

\title{GRAIN SURVIVAL IN SUPERNOVA REMNANTS AND HERBIG-HARO OBJECTS}

\author{HIDEAKI MOURI}
\affil{Meteorological Research Institute, Nagamine 1-1, Tsukuba 305-0052, Japan; hmouri@mri-jma.go.jp}

\and

\author{YOSHIAKI TANIGUCHI}
\affil{Astronomical Institute, School of Science, Tohoku University, Aoba, Sendai 980-8578, Japan}

\begin{abstract}

By using the flux ratio [Fe\,II] $\lambda$8617/[O\,I] $\lambda$6300, we demonstrate that most of the interstellar dust grains survive in shocks associated with supernova remnants and Herbig-Haro objects. The [Fe\,II]/[O\,I] flux ratio is sensitive to the gas-phase Fe/O abundance ratio, but is insensitive to the ionization state, temperature, and density of the gas. We calculate the [Fe\,II]/[O\,I] flux ratio in shocks, and compare the results with the observational data. When only 20\% of iron is in the gas phase, the models reproduce most successfully the observations. This finding is in conflict with the current consensus that shocks destroy almost all the grains and $\sim 100$\% of metals are in the gas phase. We comment on previous works on grain destruction, and discuss why grains are not destroyed in shocks.

\end{abstract}

\keywords{dust, extinction --- 
          ISM: abundances  --- 
          shock waves ---
          supernova remnants}

\section{INTRODUCTION}
For interstellar dust grains, the predominant destruction process is shocks driven by supernova explosions. In the postshock cooling gas, a charged grain is accelerated around the magnetic field line (betatron acceleration), collides with other grains and gas particles, and thereby loses its mass. Generally, it is believed that almost all the grains are destroyed in a single shock. The references are summarized in Savage \& Sembach (1996) and Jones (2000). However, we would like to argue that the actual efficiency of grain destruction is as low as $\sim 20$\% by mass in representative shock-heated nebulae, i.e., supernova remnants (SNRs) and Herbig-Haro (HH) objects.

The relative intensity of the emission lines [Fe\,II] $\lambda$8617 and [O\,I] $\lambda$6300 is used to estimate the gas-phase Fe/O abundance ratio. In the usual interstellar medium, iron is depleted into grains by a factor of $\ge$ 100 as a major dust constituent, while oxygen is largely undepleted (Savage \& Sembach 1996). Thus the gas-phase Fe/O ratio is proportional to the mass fraction of destroyed grains. The [Fe\,II]/[O\,I] flux ratio is sensitive to the gas-phase Fe/O ratio, but is insensitive to the ionization state, temperature, and density of the gas. This is because the same physical conditions are required to generate the [Fe\,II] and [O\,I] emissions. They are excited by electron collisions. Since the ionization potentials of Fe$^+$ and O$^0$ are only 16.2 and 13.6 eV, both the [Fe\,II] and [O\,I] emissions are generated in a partially ionized zone. The excitation energies of the [Fe\,II] and [O\,I] lines are 19,000 and 23,000 K. Their critical densities for collisional de-excitation at $10^4$ K are 3.5 $\times$ $10^5$ and 1.8 $\times$ $10^6$ cm$^{-3}$, which are well above the typical electron density in shocks. Moreover, the [Fe\,II] and [O\,I] lines are prominent in shocks. The grain destruction is expected to have been completed in their emission region, which is far downstream from the shock front.

We calculate the [Fe\,II]/[O\,I] flux ratio in shocks, and compare the results with the observational data of SNRs and HH objects. The analysis and subsequent discussion employ the same values for atomic constants and interstellar abundances, the references of which are given below.

\section{OBSERVATIONAL DATA}
Figure 1 shows the number distribution of the flux ratio [Fe\,II] $\lambda$8617/[O\,I] $\lambda$6300 in SNRs ({\it filled areas}) and HH objects ({\it open areas}) in our Galaxy and Magellanic Clouds. We do not include young SNRs, where supernova ejecta dominate the line emitting gas. The total (gas $+$ dust) abundances of metals in our sample are hence equal to those of the usual interstellar medium. The possible scatter of the total abundances among objects would be too small to affect the present analysis. For reference, we also show the flux ratios in H\,II regions M8 and M42 ({\it arrows}). The data were taken from the literature and are described in the figure caption.

%\placefigure{Fig.1}

The distributions of SNRs and HH objects seem to be the same. Their [Fe\,II]/[O\,I] flux ratios are higher than those of H\,II regions by factors of 2--3. Since the gas-phase fraction of iron in H\,II regions is 5--10\% (Baldwin et al. 1996; Esteban et al. 1999), the grain destruction efficiency in SNRs and HH objects seems to be 20--30\%. This result is confirmed by the following numerical calculation.

\section{NUMERICAL CALCULATION}

Our numerical calculation was based on the code MAPPINGS III, version 1.0.0g (Dopita \& Sutherland 1996). To study [Fe\,II] and [O\,I] emissions in detail, we included the charge exchange reaction Fe$^{2+}$ $+$ H$^0$ $\leftrightarrow$ Fe$^+$ $+$ H$^+$ (Neufeld \& Dalgarno 1987), updated the collision strengths of Fe$^+$ and O$^0$  with the values of Pradhan \& Zhang (1993) and Berrington \& Burke (1981), and updated the radiative transition probabilities of O$^0$  with the values in Osterbrock (1989). Careful analytic fits were made to the temperature dependence of those collision strengths.

The most important parameter for our calculation is the gas-phase elemental abundances. We included 11 elements: H, He, C, N, O, Ne, Mg, Si, S, Ar, and Fe. Formerly, the solar values had been used for the total (gas $+$ dust) abundances of metals in the interstellar medium. Recently, however, studies of elemental compositions in nearby stars revealed that the Sun is enhanced anomalously in metallicity by a factor of $\sim 1.5$ (Snow \& Witt 1996). Since no reliable data of the interstellar abundances are currently available, we used the solar values of Anders \& Grevesse (1989) with the abundances of metals being lowered by 0.20 dex (Savage \& Sembach 1996). The gas-phase fraction of iron $\delta_{\rm Fe}$ = $n_{\rm Fe}$(gas)/$n_{\rm Fe}$(gas$+$dust) was set to be 0.1, 0.2, 0.3, or 1.0. We accordingly changed the gas-phase fractions of C, O, Mg, and Si, by assuming that the grain composition is always equal to that observed toward a reddened star $\zeta$ Oph (Savage \& Sembach 1996). When $\delta_{\rm Fe}$ = 0.2, for example, 68\% of C, 64\% of O, 22\% of Mg, and 26\% of Si are in the gas phase.

The other parameters are the shock velocity $v_s$, the preshock hydrogen nucleon density $n_{{\rm H},0}$, and the preshock magnetic field $B_0$. We set $v_s$ = 50--150 km s$^{-1}$, $n_{{\rm H},0}$ = 10 or 100 cm$^{-3}$, and $B_0$ = 3 $\mu$G. These parameter values are typical of radiative shocks in SNRs and HH objects: $v_s$ $\ge$ 100 km s$^{-1}$ in SNRs and $v_s$ $\le$ 100 km s$^{-1}$ in HH objects (Russell \& Dopita 1990).

We assumed a plane-parallel geometry and a steady flow. The preshock ionization state was determined in an iterative manner. Since the ionized zone in the preshock gas is practically absent at $v_s$ $\le$ 150 km s$^{-1}$ (Dopita \& Sutherland 1996), we ignored the contribution of the preshock gas to the emergent spectrum. The calculation was terminated when the ionized hydrogen fraction $n_{{\rm H}^+}/n_{\rm H}$ fell below $\sim 0.01$. Beyond this point, the gas becomes too cool and neutral to excite the [Fe\,II] and [O\,I] emissions. We also ignored grain opacity, heating, and cooling. Their natures in shock-heated gas are quite uncertain. They should be nonetheless unimportant to the [Fe\,II] and [O\,I] excitations (see Shields \& Kennicutt 1995).

%\placefigure{Fig.2}

Figure 2 shows the cloud structure for $\delta_{\rm Fe}$ = 0.2, $v_s$ = 100 km s$^{-1}$, and $n_{{\rm H},0}$ = 10 cm$^{-3}$ as a function of the hydrogen nucleon column density from the shock front: ($a$) temperature and densities, ($b$) ionization fractions, and ($c$) line emissivities per hydrogen nucleon, which are normalized by their peak values. These quantities vary markedly across the postshock gas. However, Fe$^+$ ions coexist with O$^0$ atoms, and the [Fe\,II] $\lambda$8617 line exhibits the same emissivity profile as the [O\,I] $\lambda$6300 line. The emissivity profile is different in other forbidden lines such as [O\,II] ($\lambda$3726 $+$ $\lambda$3729) and [O\,III] $\lambda$5007. We thereby confirm our expectation that the [Fe\,II] and [O\,I] lines are generated under the same physical conditions.

%\placefigure{Fig.3}

Figure 3 shows the flux ratio [Fe\,II] $\lambda$8617/[O\,I] $\lambda$6300 for $n_{{\rm H},0}$ = 10 cm$^{-3}$ ({\it filled circles}) and 100 cm$^{-3}$ ({\it open circles}) as a function of the shock velocity. The [Fe\,II]/[O\,I] ratio at $v_s$ $\ge$ 70 km s$^{-1}$ depends only on the gas-phase iron fraction $\delta_{\rm Fe}$. Though this is not the case at $v_s$ $<$ 70 km s$^{-1}$, such slow shocks are unimportant. They do not generate the [O\,III] $\lambda$5007 emission, which is observed in all of our SNRs and HH objects. We also show the median and maximum values of the observed [Fe\,II]/[O\,I] ratios ({\it arrows}). These values are reproduced by the models for $v_s$ $\ge$ 70 km s$^{-1}$ with $\delta_{\rm Fe}$ = 0.2 and 0.3, respectively. The preshock gas has $\delta_{\rm Fe}$ $\simeq$ 0. Hence, as suggested from the comparison with H\,II regions (Fig. 1), the grain destruction efficiency is 20\% on average and 30\% at most in radiative shocks associated with SNRs and HH objects.\footnote{
Even if the individual objects have ranges of shock parameters and the observed [Fe\,II] and [O\,I] emissions originate preferentially in slow shocks driven into dense gas, our conclusion is qualitatively correct. The median value of the observed [Fe\,II]/[O\,I] ratios is reproduced by the models for $v_s$ = 50 km s$^{-1}$ with $\delta_{\rm Fe}$ = 0.3.}

\section{DISCUSSION}
Though shocks destroy grains in SNRs and HH objects, the destruction is far from complete. Typically, 80\% of iron is still locked into grains. However, many observational studies of shock-heated nebulae conclude that the grain destruction is almost complete, as reviewed by Savage \& Sembach (1996) and Jones (2000; see also references for the data used in Fig. 1). In the usual interstellar gas, heavy metals such as Fe and Ca are depleted by factors of $10^2$--$10^4$ (Savage \& Sembach 1996). If only a small fraction of the grains is destroyed, emission and absorption lines of those metals are greatly enhanced (Fesen \& Kirshner 1980). The observer is easily misled to consider that a large fraction of the grains is destroyed. Moreover, owing to wide variations of physical quantities across the gas (Fig. 2), it is generally difficult to determine elemental abundances in shocks.

Nevertheless, conclusions similar to ours were obtained in some of the past observations of SNRs. Phillips \& Gondhalekar (1983) and Jenkins et al. (1998) observed ultraviolet absorption lines of stars behind S147 and Vela SNR, estimated column densities of gas-phase ions across these SNRs, and found depletion of Al. Raymond et al. (1988, 1997) observed ultraviolet and optical emission lines of Cygnus Loop and Vela SNR, compared their relative strengths with predictions of shock models, and found depletions of Fe and of C and Si. Oliva, Moorwood, \& Danziger (1989) observed near-infrared emission lines of RCW~103, compared their strengths with model predictions, and found depletion of Fe. Reach \& Rho (1996) detected continuum emission from grains in the far-infrared spectrum of W44. It should be noted that our result is more reliable than these previous ones. The Fe/O abundance ratio estimated from the [Fe\,II] and [O\,I] fluxes is robust with respect to the shock velocity and preshock density (Fig. 3).

Since we studied the Fe/O abundance ratio alone, the grain survival probability estimated here is applicable, strictly speaking, only to Fe-bearing grains. There could exist several types of grains and subgrains which have different survival abilities. Of importance is a careful analysis of emission and absorption lines of the other elements. The present conclusion is nonetheless general. Observations of various Galactic interstellar clouds indicate that Mn, Cr, Ni, and Ti always have the same dust-phase fraction as Fe (Savage \& Sembach 1996; Jones 2000). Though the major dust constituents Mg and Si appear to be more easily liberated to the gas phase than Fe, large fractions of Mg- and Si-bearing grains survive in shocks. The above observations indicate that, when 80\% of Fe is locked into grains, $\sim 50$\% of Mg and Si are in the dust phase.\footnote{
These dust-phase fractions were adapted from Savage \& Sembach (1996). Their gas-phase Mg abundance was scaled by a factor of 2, in order to allow for the revised Mg$^+$ oscillator strengths (Fitzpatrick 1997). Since Savage \& Sembach (1996) used the Zn abundance to normalize the Fe, Mg, and Si abundances, we made no correction for the difference in the assumed total (gas $+$ dust) abundances. Noticeably, in Galactic clouds, the observed $\delta_{\rm Fe}$ value is always less than 0.3. This fact supports our conclusion for SNRs and HH objects.}

The present conclusion is, at least quantitatively, consistent with theoretical models. Jones, Tielens, \& Hollenbach (1996) predicted that 60\% (by mass) of silicate grains survive in a shock with $v_s$ = 100 km s$^{-1}$, $n_{{\rm H},0}$ = 25 cm$^{-3}$, and $B_0$ = 3 $\mu$G. This predicted probability of grain survival is somewhat low, but we could increase it. The above model assumes that grains are solid and homogeneous. If the grains are porous, e.g., consisting of several types and sizes of subgrains, they undergo less destruction (Jones et al. 1994). This is because their effective cross section is large. The resultant large gas drug prevents efficient betatron acceleration. Such porous grains are the natural result of coagulation of small grains into larger ones, and are found as interplanetary dust particles. The presence of porous grains is also suggested by the recent finding that the Sun is overabundant in heavy elements (Snow \& Witt 1996). The amount of metals available for grains in the interstellar space is much less than that had been estimated from the solar abundances. However, the observed interstellar extinction per unit length puts a lower limit to the volume fraction of space occupied by grains. This situation calls for porous grains which have high volume-to-mass ratios (see also Jones et al. 1996; Mathis 1990, 1998).

The destruction efficiency in shocks determines lifetime of interstellar grains. We estimate the grain lifetime in our Galaxy (Tielens 1998). The gas mass shocked by a supernova to a velocity equal to or greater than $v_s$ is 2500 ($v_s$/100 km s$^{-1}$)$^{-9/7}$ $M_\sun$. Since the effective supernova rate is 8 $\times$ 10$^{-3}$ yr$^{-1}$ and the mass of diffuse gas is 5 $\times$ $10^8$ $M_\sun$, the time interval for a grain to experience a supernova-driven shock with $v_s$ $\ge$ 100 km s$^{-1}$ is 3 $\times$ 10$^7$ yr. If each of the shocks destroys 20\% of the grains, their mean lifetime is 2 $\times$ 10$^8$ yr. The lifetime of the gas parcel itself is much longer, i.e., 2 $\times$ 10$^9$ yr, which is estimated from the total gas mass 8 $\times$ 10$^9$ $M_\sun$ and the star formation rate 5 $M_\sun$ yr$^{-1}$. Since interstellar grains are actually present, there has to exist some growth process, e.g., accretion of gas particles onto grains in dense gas (Jones et al. 1994). Tielens (1998) obtained a similar grain lifetime, from the metal depletions observed in diffuse and dense gases and the timescale for cycling the material between them.

Finally, we underline that dust depletion is crucial to understanding spectra of shock-heated nebulae. Their gas-phase abundances are often assumed to represent the total (gas $+$ dust) abundances of the preshock gas. This assumption could be wrong. For example, near-infrared [Fe\,II] emission lines at 1.257 and 1.644 $\mu$m are more prominent by factors of $\sim 500$ in SNRs than in H\,II regions (Graham, Wright, \& Longmore 1987). This fact is often explained by shock destruction of Fe-bearing grains. However, Mouri, Kawara, \& Taniguchi (2000) found with the code MAPPINGS III that the flux ratio [Fe\,II] 1.257 $\mu$m/Pa$\beta$ observed in SNRs is reproduced only when the gas-phase iron abundance is as low as in the H\,II region M42. The flux ratio predicted for the solar abundance is too high. This finding motivated the present work. We adopted the flux ratio [Fe\,II] $\lambda$8617/[O\,I] $\lambda$6300 as a more reliable diagnostic, conducted numerical calculations in more detail for gas-phase metallicity, and thereby determined more precisely the gas-phase iron abundance.

\acknowledgments
The authors are grateful to R. S. Sutherland for making the excellent code MAPPINGS III available to the public, and also to K. Kawara for interesting discussion.

\clearpage

\figcaption[Fig1]{Number distribution of the flux ratio [Fe\,II] $\lambda$8617/[O\,I] $\lambda$6300. Filled areas denote SNRs: RCW 86, RCW 103, N19, N49, N63A, N103B, and 0104$-$723 (Dennefeld 1986; Russell \& Dopita 1990; Vancura et al. 1992). Open areas denote HH objects: HH 1, HH 2, HH 32, HH 43, HH 56, HH 101, and Burnham's Nebula (Brugel et al. 1981; Solf et al. 1988; B\"ohm \& Solf 1990; Russell \& Dopita 1990). Arrows indicate the flux ratios in H\,II regions M8 and M42 (Osterbrock et al. 1992; Esteban et al. 1999). We independently used observations of different positions in the same object. The reddening was corrected with the reddening law of Mathis (1990) and the flux ratio H$\alpha$/H$\beta$. In the SNRs and HH objects, the intrinsic H$\alpha$/H$\beta$ ratio was assumed to be 3.00, a typical value in shocks (Fesen \& Kirshner 1980). In the H\,II regions, the ratio was assumed to be 2.85, the Case B recombination value for $T_e$ = $10^4$ K and $n_e$ = $10^4$ cm$^{-3}$ (Osterbrock 1989). This difference in the intrinsic H$\alpha$/H$\beta$ ratio is unimportant. Even if the H$\alpha$/H$\beta$ ratio were the same, the dereddened [Fe\,II]/[O\,I] ratios in the SNRs and HH objects would be simply less enhanced over those in the H\,II regions. The [Fe\,II] $\lambda$8617 line in the SNRs and HH objects is not resolved from the Pa14 line at 8598 \AA. This contamination is unimportant. Their [Fe\,II] line is much more prominent than such a weak H\,I line. In the H\,II regions, the [Fe\,II] line is resolved from the Pa14 line. \label{Fig.1}}

\figcaption[Fig2]{Cloud structure for $\delta_{\rm Fe}$ = 0.2, $v_s$ = 100 km s$^{-1}$, and $n_{{\rm H},0}$ = 10 cm$^{-3}$. The abscissa is the hydrogen nucleon column density from the shock front. ($a$) Electron temperature $T_e$, hydrogen nucleon density $n_{\rm H}$, and electron density $n_e$. ($b$) Ionization fractions of Fe$^+$, H$^+$, O$^0$, O$^+$, and O$^{2+}$. ($c$) Line emissivities per hydrogen nucleon of [Fe\,II] $\lambda$8617, [O\,I] $\lambda$6300, [O\,II] ($\lambda 3726 + \lambda 3729$), and [O\,III] $\lambda$5007. These emissivities are normalized by their peak values. \label{Fig.2}}

\figcaption[Fig3]{Numerical predictions for the flux ratio [Fe\,II] $\lambda$8617/[O\,I] $\lambda$6300. The abscissa is the shock velocity $v_s$. The gas-phase iron fraction $\delta_{\rm Fe}$ is 0.1, 0.2, 0.3, or 1.0. Filled circles denote $n_{{\rm H},0}$ = 10 cm$^{-3}$. Open circles denote $n_{{\rm H},0}$ = 100 cm$^{-3}$. Arrows indicate the median and maximum values of the observational data in Fig. 1. \label{Fig.3}}

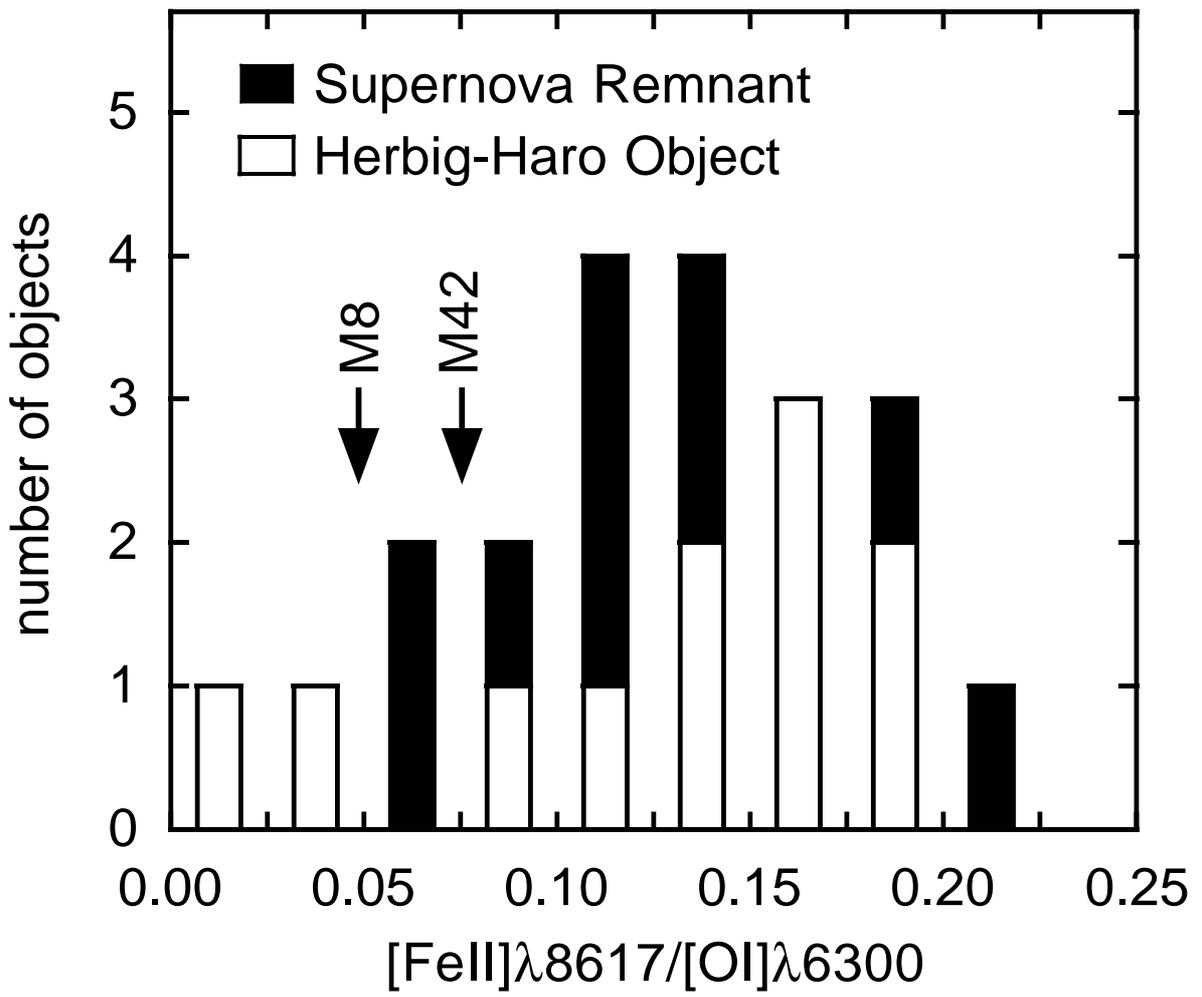
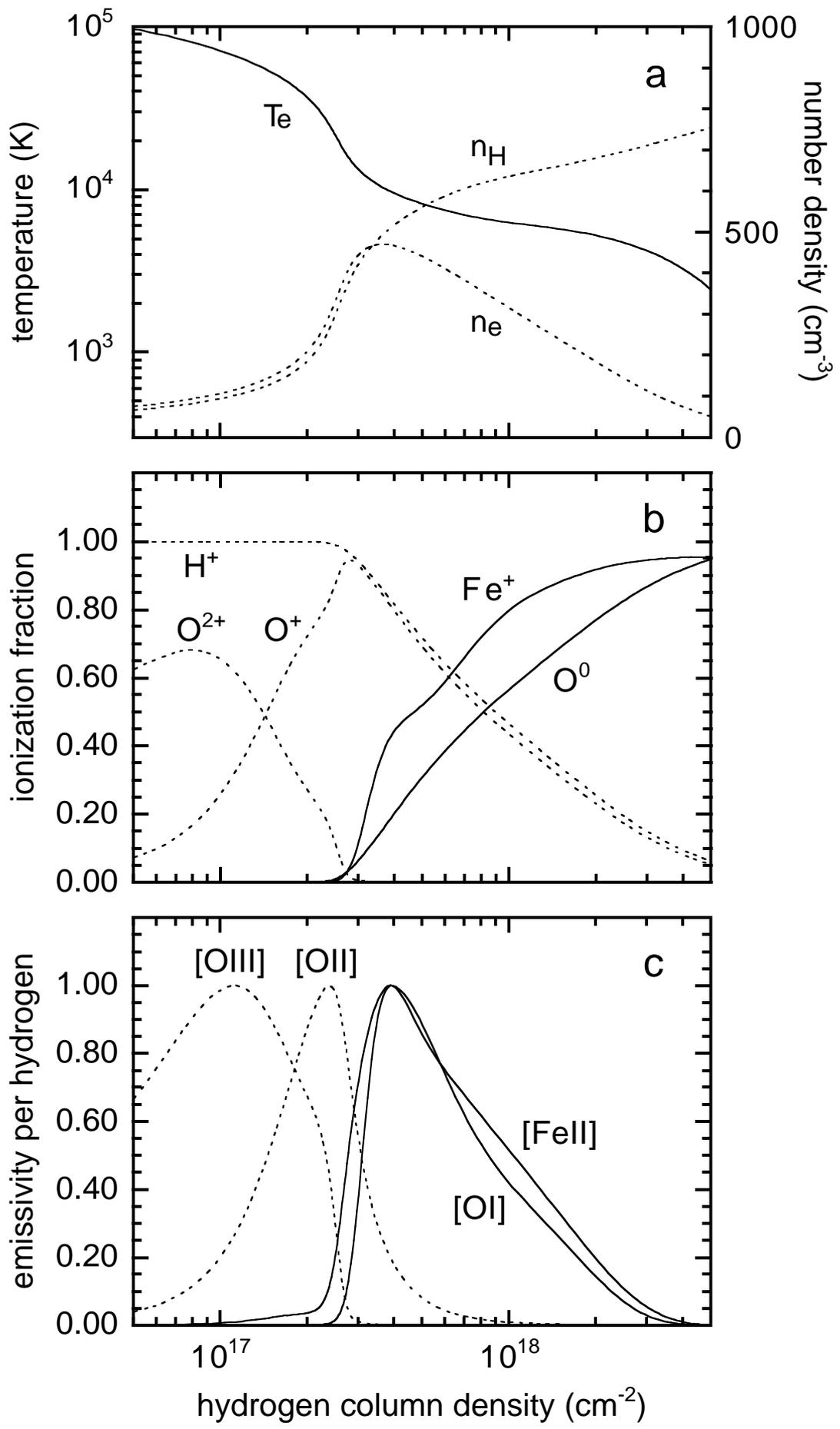
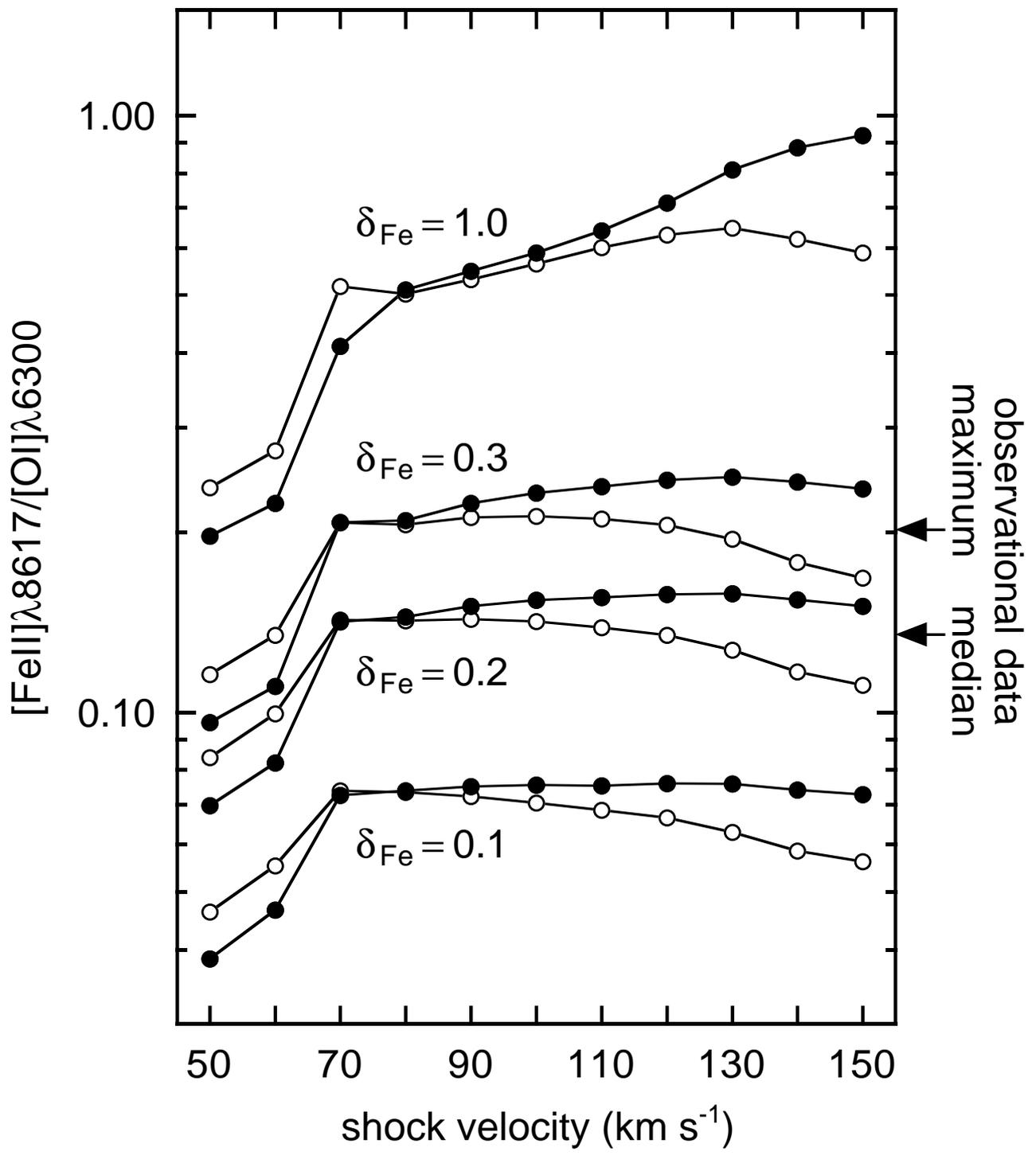
\end{document}